\documentclass[twocolumn,superscriptaddress, aps,pra ]{revtex4-1}
\makeatletter
\newcommand*{\rom}[1]{\expandafter\@slowromancap\romannumeral #1@}
\makeatother
\usepackage{color}
\usepackage{graphicx}
\usepackage{hyperref}
\usepackage{amsmath,amssymb}
\usepackage{epstopdf}
\usepackage{subcaption}
\graphicspath{{figs/}}
\begin{document}
\title{Entanglement Fidelity Ratio for Elastic Collisions in Non-Ideal Two-Temperature Dense Plasma}

\author{Ramin Roozehdar Mogaddam}
\affiliation{Department of Physics, Ferdowsi University of Mashhad,   P.O. Box 1436 Mashhad, Iran}
\author {Nasser Sepehri Javan}
\email{sepehri${_}$javan@uma.ac.ir}
\affiliation{Department of Physics, University of Mohaghegh Ardabili, P.O. Box 179, Ardabil, Iran}

\author{Kourosh Javidan}
\affiliation{Department of Physics, Ferdowsi University of Mashhad,   P.O. Box 1436 Mashhad, Iran}

\author{Hosein Mohammadzadeh}
\affiliation{Department of Physics, University of Mohaghegh Ardabili, P.O. Box 179, Ardabil, Iran}

\date{\today}

\begin{abstract}
  The quantum diffraction and symmetry effects on the entanglement fidelity (EF) of different elastic electron-electron, ion-ion and electron-ion interactions are investigated in non-ideal dense plasma. The partial wave analysis and an effective screened interaction potential including quantum mechanical diffraction and symmetry effects are employed to obtain the EF in a non-ideal dense plasma. We show that collision energy and temperatures of electron and ion have a destroying role in the entanglement. In fact, by decreasing the temperature of any kind of particles, the quantum effects become dominant and the entanglement grows up. Also, increase in the density of plasma leads to the enhancement of entanglement ratio.
\end{abstract}

\pacs{05.40.-a, 45.70.Cc, 11.25.Hf, 05.45.Df}
\keywords{Dense Plasma, Effective Potential, Entanglement Ratio}

\maketitle

\section{Introduction}\label{1}
Investigations of entanglement variations have been developed in various areas of physics. Entanglement generation due to the expansion of universe and Lorentz invariance violation\cite{ball2006entanglement,fuentes2010entanglement,friis2013entanglement,mohammadzadeh2015entanglement,farahmand2017quantum,mohammadzadeh2017entropy}, entanglement degradation because of the acceleration of observers \cite{fuentes2005alice,downes2011entangling,mehri2011pseudo,farahmand2017residual}, variations of entanglement resulted from the environmental interactions \cite{zyczkowski2001dynamics,nha2004entanglement,schliemann2003electron} are some examples in this research field. In fact, the entanglement between particles can be affected by every interacting environment, dynamical background or environmental noises. Plasma is a highly interacting environment because of the high density of charged particles which can lead to the appearance of collective phenomena via long-range electromagnetic interactions. In quantum plasmas, the main task in the study of entanglement between an incident particle and another selected species of plasma is the definition of an appropriate potential which should include all important aspects of interactions.
The study of elastic collision in plasmas has recently attracted an increasing interest since this process can reveal suitable pieces of information about plasma parameters which may have essential role in the plasma diagnostic tools design \cite{shevelko2013atomic,beyer2016introduction,fujimoto2008plasma,ramazanov2002effective,ramazanov2003scattering,ramazanov2005effective,ramazanov2005runaway}. One of the first simplest mechanisms describing interaction of charged particles in plasma is the Debye-Huckel screened potential which is known as the ideal plasma model where the interaction energy between particles is small or comparable with the average kinetic energy of a particle \cite{kvasnica1973bremsstrahlung,ramazanov2001coulomb}. This model can be idealistic for dilute plasmas, however, by increasing the plasma density, in the so-called non-ideal plasma model, multi-particle correlations originated from the simultaneous interactions of multiple charged particles should be considered in the effective potential \cite{kvasnica1975bremsstrahlung}. In this approach, avoiding to define the effective potential by the conventional Debye-Huckel model which is constructed through classical Boltzmann distribution for charged particles, the quantum mechanical diffraction and symmetry effects resulted from the collective plasma interactions are considered \cite{ramazanov2015effective}. The average separation of particles in the dense plasma is in the order of or less than the thermal de Broglie wavelength of particles which causes a high probability for the inter-particle collisions in a short distance. Therefore, it is better to consider wave aspect for the colliding particles taking into account some quantum-mechanical effects such as diffraction and symmetry \cite{ramazanov2015effective,filinov2004temperature,deutsch1981nodal}. Such quantum mechanical aspect of plasma has the vast applications in various fields including intense laser-solid density plasma interaction \cite{kremp1999quantum,marklund2006nonlinear,lower1994uniform,roozehdar2018perturbative,roozehdar2019modulation}, ion accelerator \cite{tahir2011generation}, dense plasmas appeared in the core of planets \cite{french2012ab}, astrophysical and cosmological environments like white dwarf stars \cite{opher2001nuclear,jung2001quantum}, so-called ultra-small electronic devices \cite{markowich1990semiconductor} and metal nanostructures \cite{manfredi2005model}.
Recently, quantum entanglement or correlation among distinct quantum systems is a crucial concept for the feasibility of some new modern cutting-edge technologies like quantum information and quantum communications \cite{xiang2007entanglement,falaye2019entanglement,akhtarshenas2015computable}. A new definition of entanglement measure in terms of wavepacket localization has been proposed by Fedorov et al. \cite{fedorov2004packet} and this essential parameter has been called entanglement fidelity (EF). The EF for scattering processes has been widely noted because it represents that the quantum correlation is an important characteristic in realization of the quantum measurement and information processing \cite{mishima2004entanglement,jung2011plasmon}. The general scattering processes from the entanglement point of view are theoretically studied in more detail in \cite{newton2013scattering}. It is clear that because of the quantum effects, the physical properties of quantum plasma are different from those in classical one which is studied by the classical regime or Debye-Huckel model.	Hence, it would be expected that the EF for the elastic $\alpha-\beta$ particle species (ion or electron) collisions in non-ideal dense hot quantum plasmas would be quite different from those in ideal plasma due to the collective interactions. Moreover, it can be also anticipated that the study on the EF for elastic collisions in non-ideal dense plasmas gives valuable information about the physical properties and characteristics of the quantum screening and plasma parameters. In the present work, we theoretically study the quantum and screening effects on the EF for the elastic $\alpha-\beta$ particles collision in non-ideal dense plasma. For short-range interactions of particles, the effect of surrounding plasma medium on the potential of particles collisions is taken into account through some quantum considerations which lead to some corrections on the classical Debye screening terms. In the effective interaction model, these quantum corrections are related to the quantum diffraction and symmetry effects.  To save the generality of problem, temperature of each sort of plasma particles is considered different that can be happened in the non-equilibrium plasma where the different species cannot reach to a thermodynamic equilibrium with each other. Also, partial wave method is employed to investigate the EF for the elastic particle collisions as a function of the thermal de Broglie wavelength and projectile energy. The EF rate is studied for different kinds of particles interaction including electron-electron, electron-ion and ion-ion collisions.
The paper is organized as follows: We summarize the evaluation of EF in a plasma in Sec. \ref{2}. Effective potential for a dens plasma is introduced in Sec. \ref{3}. Two different regimes are recognizable with different effective potentials. We focus on deriving analytical relationships and evaluate the EF in Sec. \ref{4} and in more detail in subsections \ref{4.1} and \ref{4.2}. Finally, we conclude the paper in Sec. \ref{5}.

	\section{Entanglement Fidelity}\label{2}
In this section, we briefly introduce an interesting phenomena in plasma which is called the entanglement and for evaluating the measure of entanglement, we define the EF. First, we consider the collision of a particle by a scattering potential.
	The stationary-state Schrodinger equation for the potential in quantum collision processes can be written as
	
	\begin{align}\label{eq1}
		\left( {{\nabla ^2} + {k^2}} \right){\psi _k}\left( {\bf{r}} \right) = \frac{{2\mu }}{{{\hbar ^2}}}V({\bf{r}}){\psi _k}\left( {\bf{r}} \right),
	\end{align}	
	where $ {\psi _k}\left( {\bf{r}} \right)$  is the solution of the scattered wave equation, $k = \sqrt {2\mu E/{\hbar ^2}} $ is the wave number, $\mu $  is the reduced mass of the collision system, $V(r)$ is scattering potential,  $E= {{\mu {v^2}}}/{2}$ is the kinetic energy of the projectile, $v$ is the collision velocity, and $\hbar$ is the rationalized Plank constant. Here the final state wave function $ {\psi _k}\left( {\bf{r}} \right)$  would be represented by the partial wave expansion \cite{smirnov2008plasma} in the following form
	\begin{align}\label{eq2}
		{\psi _k}\left( {\bf{r}} \right) = \sum\limits_{l = 0}^\infty  {{i^l}\left( {2l + 1} \right)} {D_l}(k){P_l}(\cos \theta ){R_l}(r),
	\end{align}	
	where, ${D_l}(k)$  is the expansion coefficient, $i$  is the pure imaginary number, ${R_l}(r)$ is the solution of the radial wave equation, ${P_l}(\cos \theta )$ is the Legendre polynomial of order $l$ , and $l$ is the angular momentum quantum number. For a spherically symmetric potential $V({\bf{r}})$, it has been shown that the radial wave equation and the expansion coefficient ${D_l}(k)$  are given by \cite{mishima2004entanglement}	
	\begin{align}\label{eq3}
		\left[ {\frac{1}{{{r^2}}}\frac{d}{{dr}}\left( {{r^2}\frac{d}{{dr}}} \right) - \frac{{l\left( {l + 1} \right)}}{{{r^2}}} - \frac{{2\mu }}{{{\hbar ^2}}}V(r) + {k^2}} \right]{R_l}(r) = 0,
	\end{align}
	\begin{align}\label{eq4}
		{D_l}(k) = {(2\pi )^{ - 3/2}}{\left[ {1 + \frac{{2i\mu k}}{{{\hbar ^2}}}\int_0^\infty  {dr\,{r^2}{j_l}(kr)V(r){R_l}(r)} } \right]^{ - 1}},
	\end{align}
respectively, where
\begin{widetext}	
 ${j_l}(kr)$  is the spherical Bessel function of order  $l$. The solution of the radial wave equation is represented by
\begin{eqnarray}\label{eq5}
		{R_l}(r) = {j_l}(kr) + \frac{{2\mu k}}{{{\hbar ^2}}}\left( {{n_l}\left( {kr} \right)\int_0^r {dr'{{r'}^2}} {j_l}(kr')V(r'){R_l}(r')
                   +{j_l}(kr)\int_r^\infty  {dr'{{r'}^2}} {n_l}(kr')V(r'){R_l}(r')}\right),
	\end{eqnarray}
where ${n_l}\left( {kr} \right)$  is the spherical Neumann function of order $l$. The asymptotic form of the radial wave function can be achieved by the phase-shift ${\delta _l}$ such as ${R_l}(r) \propto {(kr)^{ - 1}}\sin (kr - \pi l/2 + {\delta _l})$ .
\end{widetext}

The entanglement generation by the scattering processes has been investigated by K. Mishima, M. Hayashi and S. H. Lin \cite{mishima2004entanglement}.
It has been shown that the collisional EF for the scattering process can be represented by ${f_k} \propto {\left| {\int {{d^3}{\bf{r}}\,} {\psi _k}\left( {\bf{r}} \right)} \right|^2}$, that is, the absolute square of the scattered wave function for a given interaction potential \cite{mishima2004entanglement}. In low collision energies, the main contribution of the collision is related to the partial $s$-wave scattering ($l = 0$). Therefore, the EF, i.e. ${f_k}$, can be stated by using the expansion coefficient ${D_l}(k)$ and the radial wave equation ${R_l}(r)$, as follows	
	\begin{align}\label{eq6}
		{f_k} \propto \frac{{{{\left| {\int_0^\infty  {dr\,{r^2}{j_0}\left( {kr} \right)} } \right|}^2}}}{{1 + {{\left| {\frac{{2\mu k}}{{{\hbar ^2}}}\int_0^\infty  {dr\,{r^2}V(r){j_0}\left( {kr} \right)} } \right|}^2}}}
	\end{align}	
	Now, the collisional EF in the low energies for elastic collisions between $\alpha$ and $\beta$; two different or the same species particles, in a plasma with an appropriate effective potential can be evaluated as follows	
	\begin{align}\label{eq7}
		{f_k} \propto \frac{{{{\left| {\int_0^\infty  {dr\,{r^2}\frac{{\sin (kr)}}{{kr}}} } \right|}^2}}}{{1 + {{\left| {\frac{{2\mu }}{{{\hbar ^2}}}\int_0^\infty  {dr\,{r^2}{\varphi _{\alpha \beta }}(r)\frac{{\sin (kr)}}{{kr}}} } \right|}^2}}},
	\end{align}
	where, ${\varphi _{\alpha \beta }}(r)$ describes the effective interaction potential between the projectile $\alpha$ and the screened $\beta$ particles.

\section{Non ideal dens plasma and effective potential}\label{3}
The effective interparticle interaction potential in plasma can be derived in two different methods. In the first method, one obtains the effective potential using the solution of generalized Poisson-Boltzmann equation \cite{arkhipov2011self}. The second one, is related to the dielectric response function \cite{gericke2010screening}. Recently, using the second method, the effective potentials of interactions of a non-ideal, non-isothermal plasma has been investigated by Ramazanov {\it et. al.} \cite{ramazanov2015effective}. The effective potential of the $\alpha  - \beta$ particle species (ion or electron) interaction in non-ideal dense hot plasma with effective screening taking into account quantum-mechanical diffraction and symmetry effects with a strongly coupled ion and semiclassical electron subsystems is given by \cite{ramazanov2015effective}

\begin{widetext}	
	\begin{align}\label{eq8}
		{\varphi _{\alpha \beta }}(r) = \frac{{{Z_\alpha }{Z_\beta }{e^2}}}{r}\frac{1}{{{\gamma ^2}\sqrt {1 - {{\left( {2{k_D}/{\lambda _{ee}}{\gamma ^2}} \right)}^2}} }}\left[ {\left( {\frac{{1/{\lambda _{ee}}^2 - {B^2}}}{{1 - {B^2}{\lambda _{\alpha \beta }}^2}}} \right){e^{ - Br}} - \left( {\frac{{1/{\lambda _{ee}}^2 - {A^2}}}{{1 - {A^2}{\lambda _{\alpha \beta }}^2}}} \right){e^{ - Ar}}} \right]\,\, - \frac{{{Z_\alpha }{Z_\beta }{e^2}}}{r}\frac{{1 - {\delta _{\alpha \beta }}}}{{1 + {C_{\alpha \beta }}}}{e^{ - r/{\lambda _{\alpha \beta }}}},
	\end{align}	
where  ${Z_\alpha }$ (${Z_\beta }$), $e$ and ${\lambda _{\alpha \beta }} = {\hbar }/{{\sqrt {4\pi {\mu _{\alpha \beta }}{k_\beta }{T_{\alpha \beta }}} }}$  are atomic numbers of particle species $\alpha$ ($\beta$), the electron charge and thermal de-Broglie wavelength of pairs of particles $\alpha$ and $\beta$, respectively. ${\mu _{\alpha \beta }} = {{{m_\alpha }{m_\beta }}}/{{({m_\alpha } + {m_\beta })}}$ is the reduced mass, ${k_B }$ is Boltzmann constant, and ${T_{\alpha \beta }} = \sqrt {{T_\alpha }{T_\beta }} $ is defined by the temperatures of species $\alpha$ ($T_{\alpha}$) and $\beta$ ($T_{\beta}$). Also, ${k_D} =\sqrt{ {k_e}^2 + {k_i}^2}$ is the screening parameter considering the contributions of electrons and ions, where ${k_e} = \sqrt {{{4\pi {n_e}{e^2}}}/{{{k_B }{T_e}}}} $, ${k_i} = \sqrt {{{4\pi {n_i}{e^2}}}/{{{k_B }{T_i}}}}$ , and  ${\gamma ^2} = {k_i}^2 + 1/{\lambda _{ee}}^2$. Also, $A$ , $B$ and ${C_{\alpha \beta }}$  have been defined as
	
	\begin{align}\label{eq9}
		&{A^2} = \frac{{{\gamma ^2}}}{2}\left( {1 + \sqrt {1 - {{\left( {\frac{{2{k_D}}}{{{\lambda _{ee}}{\gamma ^2}}}} \right)}^2}} } \right),\\
		&{B^2} = \frac{{{\gamma ^2}}}{2}\left( {1 - \sqrt {1 - {{\left( {\frac{{2{k_D}}}{{{\lambda _{ee}}{\gamma ^2}}}} \right)}^2}} } \right),\\
		&{C_{\alpha \beta }} = \frac{{{k_D}^2{\lambda _{\alpha \beta }}^2 - {k_i}^2{\lambda _{ee}}^2}}{{{\lambda _{ee}}^2/{\lambda _{\alpha \beta }}^2 - 1}}.
	\end{align}
It is worth mentioning that the effective potential in Eq. (\ref{eq8}) is valid for ${\left( {2{k_D}/{\lambda _{ee}}{\gamma ^2}} \right)^2} < 1$.	
	For the case ${\left( {2{k_D}/{\lambda _{ee}}{\gamma ^2}} \right)^2} > 1$, the effective potential takes the following form \cite{ramazanov2015effective}
	
	\begin{align}\label{eq10}
		\begin{array}{l}
			{\varphi _{\alpha \beta }}(r) = \frac{{{Z_\alpha }{Z_\beta }{e^2}}}{r}\frac{{{d_{\alpha \beta }}}}{{{\gamma ^2}\sqrt {{{\left( {2{k_D}/{\lambda _{ee}}{\gamma ^2}} \right)}^2} - 1} }}\sin (\sqrt {{k_D}/{\lambda _{ee}}} \sin (\omega /2)r + {\theta _{\alpha \beta }})\exp \left[ { - r\sqrt {{k_D}/{\lambda _{ee}}} \cos (\omega /2)} \right]\\
			\,\,\,\,\,\,\,\,\,\,\,\,\,\,\,\,\,~~~ - \frac{{{Z_\alpha }{Z_\beta }{e^2}}}{r}\frac{{1 - {\delta _{\alpha \beta }}}}{{1 + {C_{\alpha \beta }}}}{e^{ - r/{\lambda _{\alpha \beta}}},}
		\end{array}
	\end{align}	
where parameters ${d_{\alpha \beta }}$, ${\theta _{\alpha \beta }}$ and $\omega $ are	
	\begin{align}\label{eq11}
		&{d_{\alpha \beta }} = \sqrt {a_{\alpha \beta }^2 + b_{\alpha \beta }^2},\\
		&{\theta _{\alpha \beta }} = \arctan \left( {\frac{{{b_{\alpha \beta }}}}{{{a_{\alpha \beta }}}}} \right),\\
		&\omega  = \arctan \left[ {\sqrt {{{\left( {2{k_D}/{\lambda _{ee}}{\gamma ^2}} \right)}^2} - 1} } \right],
	\end{align}
${a_{\alpha \beta }}$ and ${b_{\alpha \beta }}$ are
	\begin{align}\label{eq12}
		&{a_{\alpha \beta }} = \frac{{2\left( {1/{\lambda _{ee}}^2 - {\gamma ^2}/2} \right)\left( {1 - {\gamma ^2}{\lambda _{\alpha \beta }}^2/2} \right) + {\gamma ^4}{\lambda _{\alpha \beta }}^2\left( {{{\left( {2{k_D}/{\lambda _{ee}}{\gamma ^2}} \right)}^2} - 1} \right)}}{{{{\left( {1 - {\gamma ^2}{\lambda _{\alpha \beta }}^2/2} \right)}^2} + {\gamma ^4}{\lambda _{\alpha \beta }}^4\left( {{{\left( {2{k_D}/{\lambda _{ee}}{\gamma ^2}} \right)}^2} - 1} \right)/4}},\\
		&{b_{\alpha \beta }} = \frac{{{\gamma ^2}\left( {1 - {\lambda _{\alpha \beta }}^2/{\lambda _{ee}}^2} \right)\sqrt {{{\left( {2{k_D}/{\lambda _{ee}}{\gamma ^2}} \right)}^2} - 1} }}{{{{\left( {1 - {\gamma ^2}{\lambda _{\alpha \beta }}^2/2} \right)}^2} + {\gamma ^4}{\lambda _{\alpha \beta }}^4\left( {{{\left( {2{k_D}/{\lambda _{ee}}{\gamma ^2}} \right)}^2} - 1} \right)/4}}.
	\end{align}
\end{widetext}	
We notice that the last term on the right hand side of the equations (\ref{eq8}) and (\ref{eq10}) disappears for ion-ion and electron-electron interactions.
\section{Entanglement fidelity of dense plasma}\label{4}
	The entanglement of an incident particle and a selected electron or ion in plasma is varied by the effective potential of plasma and can be quantified by the EF. Therefor, the EF is dependent on the effective potential. One can consider the relative EF by evaluation of the ratio of the EF for the effective interaction potential ${\varphi _{\alpha \beta }}$ with respect to EF of the pure coulomb potential ${V_C}\left( r \right) =  - \frac{{Z{e^2}}}{r}$  as follows
\begin{widetext}	
	\begin{align}\label{eq14}
		{R_{\alpha-\beta }}\left( {k,{\lambda _{\alpha \beta }},{n_e},{n_i}} \right) =\frac{{{f_k}^\varphi \left( {k,{\lambda _{\alpha \beta }},{n_e},{n_i}} \right)}}{{{f_k}^{Coul}\left( k \right)}} = \frac{{1 + {{\left| { - \frac{{2Z{e^2}\mu k}}{{{\hbar ^2}}}\int_0^\infty  {dr\,r\frac{{\sin (kr)}}{{kr}}} } \right|}^2}}}{{1 + {{\left| {\frac{{2\mu k}}{{{\hbar ^2}}}\int_0^\infty  {dr\,{r^2}{\varphi _{\alpha \beta }}(r)\frac{{\sin (kr)}}{{kr}}} } \right|}^2}}}.
	\end{align}
Using Eq.(\ref{eq8}), we can evaluate the entanglement fidelity ratio (EFR) for electron-electron interaction. In order to derive the evaluating the integrals, we use the following integral relation
 \begin{align}\label{eq15}
		\int_{0}^{\infty}\exp(-Cr)\sin(kr)dr=\frac{k}{k^2 + C^2}.
\end{align}
Therefor, we obtain $R_{e-e}$ (EFR for electron-electron interaction) as follows	
	\begin{align}\label{eq16}
		{R_{e - e}}\left( {\bar E,{n_e},{T_e},{n_i},{T_i}} \right) =
		\frac{{1 + \frac{4}{{\bar E}}}}{{1 + \frac{4}{{\bar E}}{{\left[ {\frac{{8\bar E/{{\bar \gamma }^2}}}{{\left( {1 + {{\bar \lambda }_{ee}}^2{{\bar k}_i}^2} \right)\left( {{{\left( {2\bar E/{{\bar \gamma }^2} + 1} \right)}^2} + 4{{\bar k}_D}^2/{{\bar \lambda }_{ee}}^2{{\bar \gamma }^4} - 1} \right)}}} \right]}^2}}}
	\end{align}
where, the dimensionless parameters are defined by
\begin{align}\label{eq17}
&\bar{E}=\frac{E}{Z^{2}R_{y}}=\frac{2E\hbar^{2}}{Z^{2}\mu^{2}e^{4}},\nonumber\\
&\bar{\gamma}=a_{z}\gamma,~~~\bar{k_{D}}=a_{z}k_{D}, \bar{\lambda_{\alpha\beta}}=\frac{\lambda_{\alpha\beta}}{a_{z}}.
\end{align}
Also, $a_z =a_0 /Z$ and $a_0 =\hbar^2 /\mu e^2$ is Bohr radius and other parameters have the conventional meaning.

In the same manner, the extraction of EFR for effective ion-ion interaction is straightforward and is given by	
	\begin{align}\label{eq18}
		{R_{i - i}}\left( {\bar E,{n_e},{T_e},{n_i},{T_i}} \right) = \frac{{1 + \frac{4}{{\bar E}}}}{{1 + \frac{4}{{\bar E}}{{\left[ {\frac{{4\left( {{{\bar E}^2}/{{\bar \gamma }^4} + \bar E/{{\bar \gamma }^4}{{\bar \lambda }_{ee}}^2} \right)}}{{\left( {{{\left( {2\bar E/{{\bar \gamma }^2} + 1} \right)}^2} + 4{{\bar k}_D}^2/{{\bar \lambda }_{ee}}^2{{\bar \gamma }^4} - 1} \right)}}} \right]}^2}}}
	\end{align}
Calculation of the EFR for electron-ion interaction is a little complicated with respect to the previous ones due to the non-vanishing term in the effective potential. In this case, we obtain
	\begin{align}\label{eq19}
		{R_{e - i}}\left( {\bar E,{n_e},{T_e},{n_i},{T_i}} \right) = \frac{{1 + \frac{4}{{\bar E}}}}{{1 + \frac{4}{{\bar E}}{F^2}}},
	\end{align}	
	where,	
	\begin{align}\label{eq20}
		\begin{array}{l}
			F = \frac{{4\left( {\left( {1/{{\bar \lambda }_{ee}}^2 - 1/{{\bar \lambda }_{ei}}^2} \right){{\bar E}^2}/{{\bar \gamma }^4} + \left( {1/{{\bar \lambda }_{ee}}^2 - 1/{{\bar \gamma }^2}{{\bar \lambda }_{ee}}^2{{\bar \lambda }_{ei}}^2 - {{\bar k}_D}^2/{{\bar \gamma }^2}{{\bar \lambda }_{ee}}^2} \right)\bar E/{{\bar \gamma }^2}} \right)}}{{\left( {1/{{\bar \lambda }_{ei}}^2 - {{\bar \gamma }^2} + {{\bar \lambda }_{ei}}^2{{\bar k}_D}^2/{{\bar \lambda }_{ee}}^2} \right)\left( {{{\left( {2\bar E/{{\bar \gamma }^2} + 1} \right)}^2} + 4{{\bar k}_D}^2/{{\bar \lambda }_{ee}}^2{{\bar \gamma }^4} - 1} \right)}}+\frac{{{{\bar \lambda }_{ei}}^2 - {{\bar \lambda }_{ee}}^2}}{{\left( {1 - {{\bar k}_i}^2{{\bar \lambda }_{ei}}^2} \right){{\bar \lambda }_{ee}}^2 - \left( {1 - {{\bar k}_D}^2{{\bar \lambda }_{ei}}^2} \right){{\bar \lambda }_{ei}}^2}}\frac{{\bar E}}{{\bar E + 1/{{\bar \lambda }_{ei}}^2}}.
		\end{array}
	\end{align}
\end{widetext}
it is better to mention that we used the effective potential which is given by Eq. (\ref{eq8}). In fact, for ${\left( {2{k_D}/{\lambda _{ee}}{\gamma ^2}} \right)^2} < 1$ all evaluations can be done analytically. For ${\left( {2{k_D}/{\lambda _{ee}}{\gamma ^2}} \right)^2} > 1$, the effective potential is given by Eq. (\ref{eq10}) and numerical calculation is needed. In current paper we focus on the analytical evaluations and the numerical calculation is postponed to the future work.

\begin{figure}
	\centerline{\includegraphics[scale=.45]{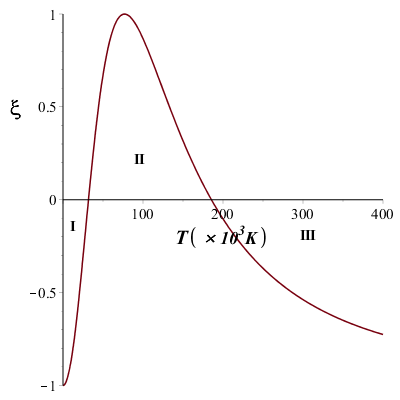}}
	\caption{The parameter $\xi={\left( {2{k_D}/{\lambda _{ee}}{\gamma ^2}} \right)^2} - 1$  as a function of the isothermal ion and electron temperature  for $n = 2 \times {10^{23}}c{m^{-3}}$. Effective potential of Eq. $(\ref{eq8})$ ( Eq. $(\ref{eq10})$) is valid in reginos $\mathrm{\rom{1}}$  and $\mathrm{\rom{3}}$ ($\mathrm{\rom{2}}$).}
	\label{fig:gg}
\end{figure}
Figure (\ref{fig:gg}), as an example, shows that for a dens and uniform temperature plasma, the effective potential of Eq. (\ref{eq8}) is valid in regions $\mathrm{\rom{1}}$ and $\mathrm{\rom{3}}$ while in region $\mathrm{\rom{2}}$ the effective potential of Eq. (\ref{eq10}) is reliable.
We consider two regions $\mathrm{\rom{1}}$ (low temperature limit) and $\mathrm{\rom{3}}$ (high temperature limit) in the following.
	\subsection{Low Temperature limit}\label{4.1}
\begin{figure}
		\begin{subfigure}{0.3\textwidth}\includegraphics[width=\textwidth]{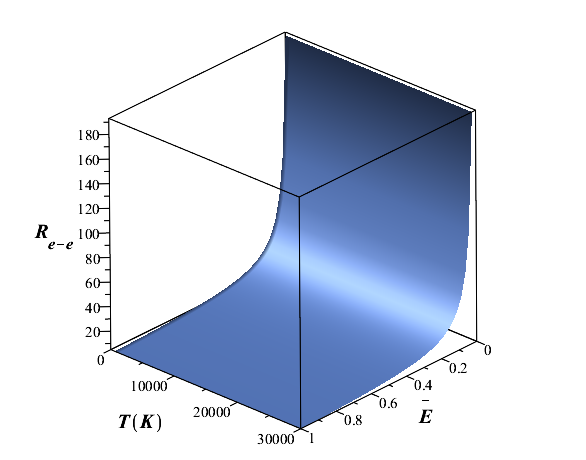}
			\caption{}
			\label{fig:2-1}
		\end{subfigure}
		\begin{subfigure}{0.3\textwidth}\includegraphics[width=\textwidth]{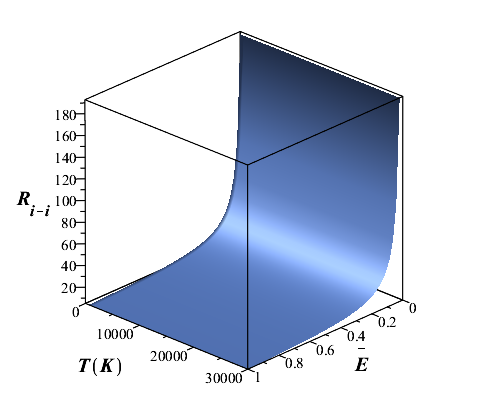}
			\caption{}
			\label{fig:2-2}
		\end{subfigure}
		\begin{subfigure}{0.36\textwidth}\includegraphics[width=\textwidth]{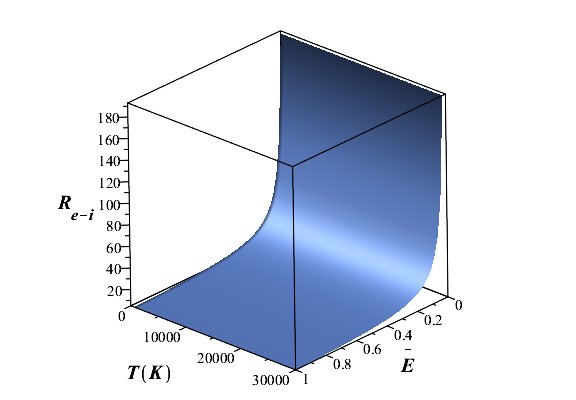}
			\caption{}
			\label{fig:2-3}
		\end{subfigure}
		\caption{EFR; ${R_{\alpha-\beta}}$, as a function of the scaled collision energy $\bar{E}$ and temperature in isothermal dens plasma for different (a) electron-electron (b) ion-ion (c) electron-ion interactions. Also, we suppose that $n_{e}=n_{i}=2\times10^{23} cm^{-3}$.}
		\label{fig:fig2}
	\end{figure}

\begin{figure}
	\centerline{\includegraphics[scale=.45]{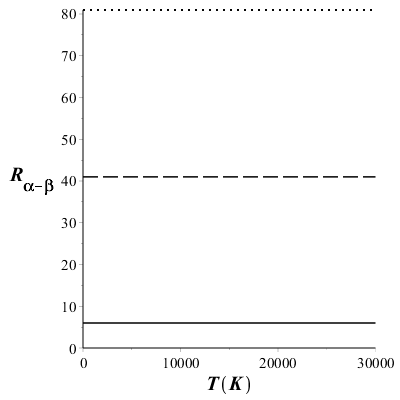}}
	\caption{EFR of all different interactions for $\bar{E}=0.8$ (solid line), $\bar{E}=0.1$ (dashed line) and $\bar{E}=0.05$ (dotted line) for dens plasma with  $n_{e}=n_{i}=2\times10^{23} cm^{-3}$ as a function of temperature.}
	\label{fig:h}
\end{figure}

For an analytical consideration, we restrict ourself to regions $\mathrm{\rom{1}}$ and $\mathrm{\rom{3}}$. In region $\mathrm{\rom{1}}$ the temperature has an upper bound. In other words, we consider the low temperature limit. EFR in this region for different electron-electron, ion-ion and ion-electron interactions is investigated diagrammatically in Fig. (\ref{fig:fig2}). It is obvious that the general behaviour is the same for all cases. EFR is an monotonically decreasing function with respect to the collision energy. We compare EFR of different interactions in Fig. (\ref{fig:h}) for different values of collision energy. One can observe that for a fixed collision energy, EFR of all interactions are independent of temperature. The effective potential for small value of collision energy is more impressive and EFR is rapidly decreased by increasing collision energy.

	\begin{figure}
	\centerline{\includegraphics[scale=.45]{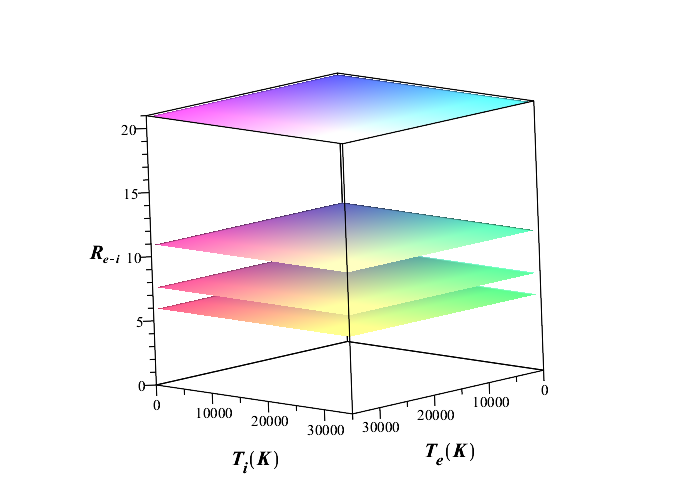}}
	\caption{EFR; ${R_{ei}}$, of two-temperature dens plasma as a function of electron and ion temperatures. Also, we suppose that $n_{e}=n_{i}=2\times10^{23} cm^{-3}$. $\bar{E}=0.1,0.2,0.6,0.8$ correspond to up to down planes, respectively.}
	\label{fig:fig3}
\end{figure}
\begin{figure}
	\begin{subfigure}{0.30\textwidth}\includegraphics[width=\textwidth]{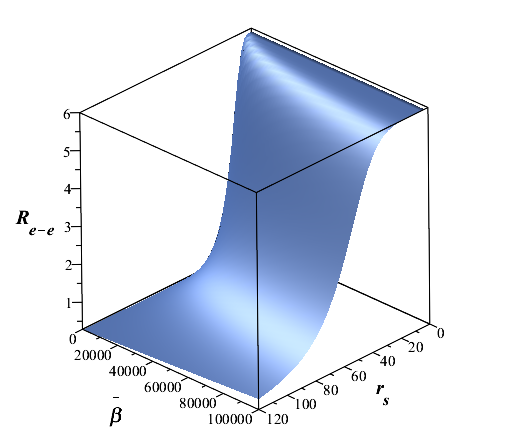}
		\caption{}
		\label{fig:4-1}
	\end{subfigure}
	\begin{subfigure}{0.30\textwidth}\includegraphics[width=\textwidth]{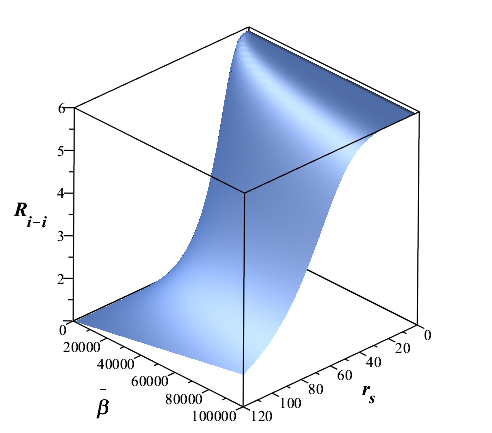}
		\caption{}
		\label{fig:4-2}
	\end{subfigure}
	\begin{subfigure}{0.30\textwidth}\includegraphics[width=\textwidth]{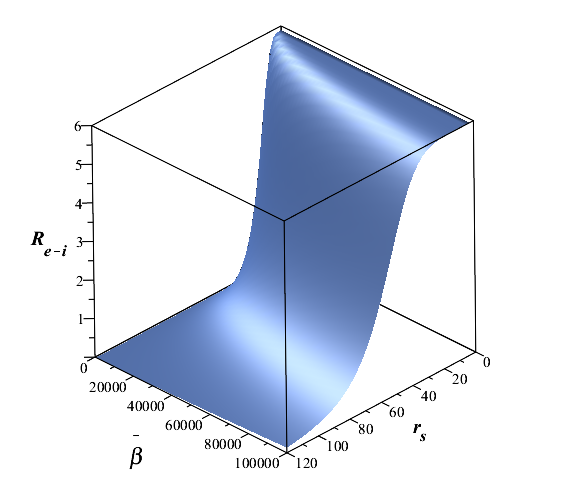}
		\caption{}
		\label{fig:4-3}
	\end{subfigure}
	\caption{EFR; ${R_{\alpha-\beta}}$, as a function of $\bar{\beta}$ and $r_s$ in isothermal dens plasma for different (a) electron-electron (b) ion-ion (c) electron-ion interactions. Also, we suppose that $n_{e}=n_{i}$ and $\bar{E}=0.8$.}
	\label{fig:fig4}
\end{figure}
\begin{figure}
	\begin{subfigure}{0.30\textwidth}\includegraphics[width=\textwidth]{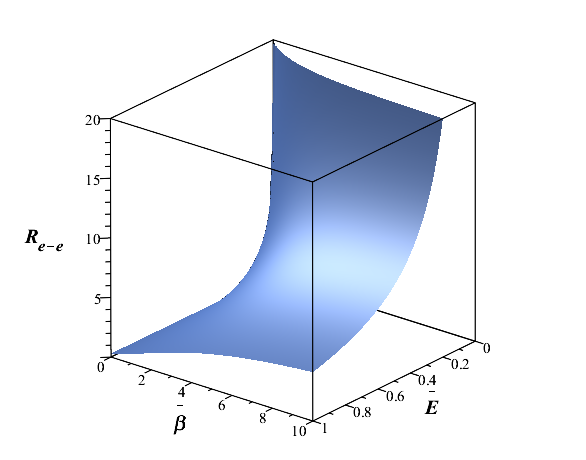}
		\caption{}
		\label{fig:5-1}
	\end{subfigure}
	\begin{subfigure}{0.30\textwidth}\includegraphics[width=\textwidth]{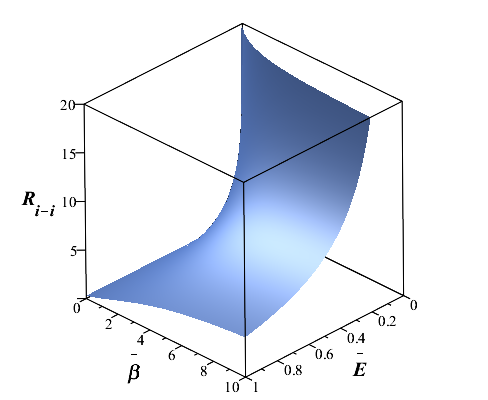}
		\caption{}
		\label{fig:5-2}
	\end{subfigure}
	\begin{subfigure}{0.360\textwidth}\includegraphics[width=\textwidth]{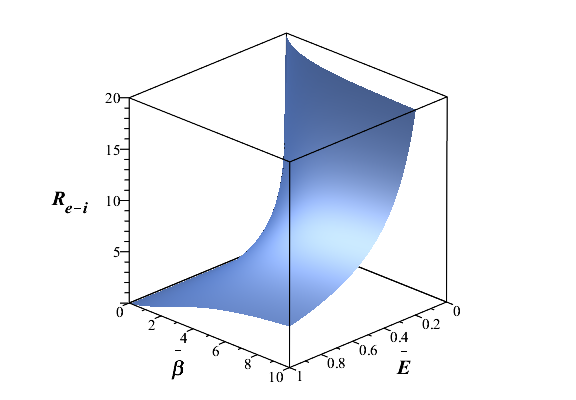}
		\caption{}
		\label{fig:5-3}
	\end{subfigure}
	\caption{EFR; ${R_{\alpha-\beta}}$, as a function of $\bar{\beta}$ and collision energy $\bar{E}$ in isothermal dens plasma for different (a) electron-electron (b) ion-ion (c) electron-ion interactions. Also, we suppose that $n_{e}=n_{i}=2\times10^{23} cm^{-3}$.}
	\label{fig:fig6}
\end{figure}
We consider non-isothermal dens plasma, in the following. For fixed values of collision energy and density of electrons and ions ($n_{e}=n_{i}=2\times10^{23} cm^{-3}$), EFR of different interactions with respect to the temperature of electrons and ions are considered diagrammatically in Fig. (\ref{fig:fig3}). Behaviour of different interactions are the same in general and we depicted only the EFR of electron-ion interaction diagrammatically. In fact, in low temperature limit, EFR does not depend on the temperatures and it is evident that by increasing the collision energy, EFR will be decreased.

	\subsection{High Temperature limit}\label{4.2}
Analytical calculation of EFR can be done in region $\mathrm{\rom{3}}$ where the temperature has the lower bound. In fact, we restrict ourself in high temperature limit.
First, we introduce a new parameter which is related to the inverse of temperature, $\bar{\beta}=1/k_BT$. Therefore, the region $\mathrm{\rom{3}}$ is identified by $0\le\bar{\beta}\le \bar{\beta}^{\star}$, where $\bar{\beta}^{\star}$ is the root of $\xi (\bar{\beta}^{\star})=0$. We notice that $\bar{\beta}=0$ corresponds to infinite temperature and $\bar{\beta}^{\star}$ is related to the inverse of lower bound temperature. Also, another dimensionless parameter which is correspond to the inverse of density is  defined as $r_s=a/a_B$. $a=(3/4\pi n_{e})^{1/3}$ is the average distance between particles and $a_B=\hbar ^2/m_e e^2$.
It is obvious that $r_s=0$ denotes the infinite density.

 Figure (\ref{fig:fig4}) shows the behavior of EFR with respect to $\bar{\beta}$  and $r_s$ for three different types of $\alpha$-$\beta$ scattering for dens plasma in high temperature limit. EFR is a monotonically increasing function of $\bar{\beta}$ while it is a decreasing function of $r_s$. It seems that  EFR vanishes in classical limit. Therefore, for high densities (small values of $r_s$) and near to the lower bound of temperature, the quantum effects are dominant and EFR grows up. Also, the general behaviour of EFR for three different interactions are the same.

\begin{figure} [h]
	\centerline{\includegraphics[scale=.4]{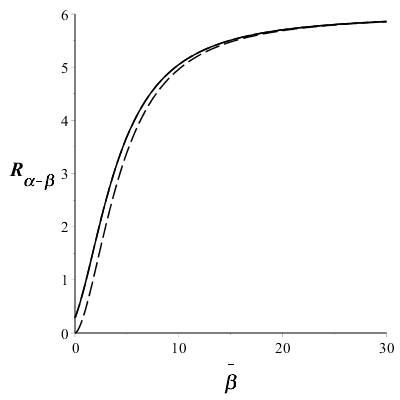}}
	\caption{EFR of electron-electron and ion-ion (solid line) and electron-ion (dashed line) iterations for dens plasma with fixed collision energy ($\bar{E}=0.8$) and $n_{e}=n_{i}=2\times10^{23} cm^{-3}$ as a function of $\bar{\beta}$.}
	\label{fig:fig5}
\end{figure}

In Fig. (\ref{fig:fig5}), a comparison between EFR of different interactions for fixed density and certain value of collision energy in isothermal plasma indicates that $R_{ii}=R_{ee}>R_{ei}$ in high temperatures. By decreasing the temperature, EFR of all interactions tends to an identical saturated value. Also, at very high temperature or classical limit, only EFR of electron-ion interaction becomes zero.

\begin{figure}[h]
	\begin{subfigure}{0.30\textwidth}\includegraphics[width=\textwidth]{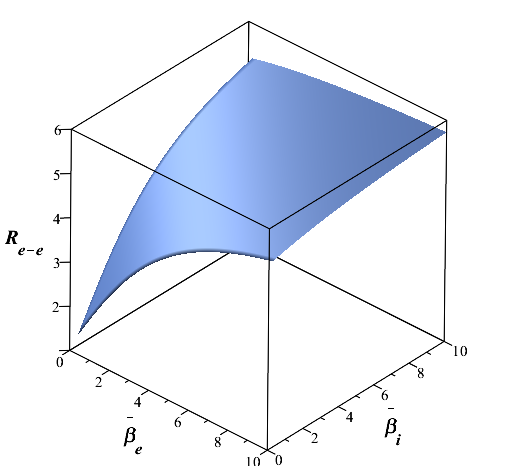}
		\caption{}
		\label{fig:6-1}
	\end{subfigure}
	\begin{subfigure}{0.30\textwidth}\includegraphics[width=\textwidth]{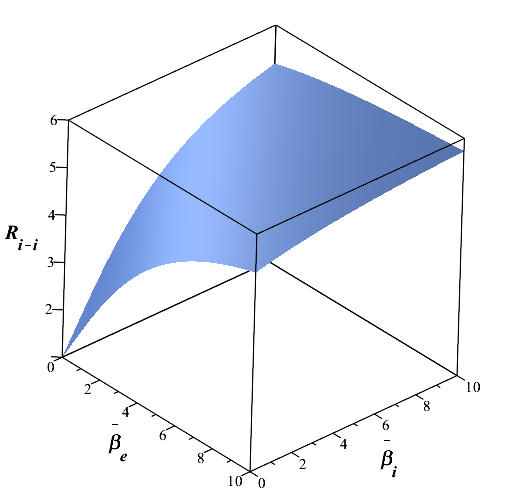}
		\caption{}
		\label{fig:6-2}
	\end{subfigure}
	\caption{EFR; ${R_{\alpha-\beta}}$, as a function of $\bar{\beta_e}$ and $\bar{\beta_i}$  in two temperature dens plasma for different (a) electron-electron (b) ion-ion interactions. Also, we suppose that $n_{e}=n_{i}=2\times10^{23} cm^{-3}$ and $\bar{E}=0.8$.}
	\label{fig:fig7}
\end{figure}

\begin{figure}
	\centerline{\includegraphics[scale=.45]{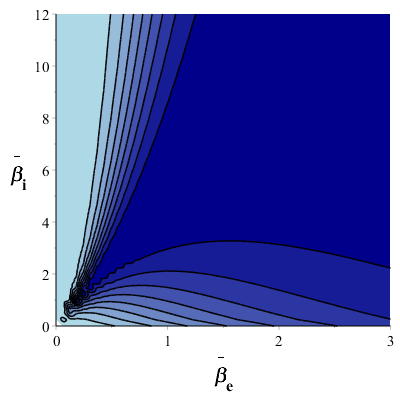}}
	\caption{Contour plot of EFR of electron-ion interaction with respect to $\bar{\beta}_e$ and $\bar{\beta}_i$ with fixed collision energy ($\bar{E}=0.8$) and $n_{e}=n_{i}=2\times10^{23} cm^{-3}$. Zero EFR corresponds to the light blue and EFR grows up by darkness of the blue color.}
	\label{fig:fig8}
\end{figure}

The variation of EFR with respect to $\bar{\beta}$ and $\bar{E}$ is depicted in Fig. (\ref{fig:fig6}). $R_{ee}$ and $R_{ii}$ have completely the same behaviour. Also, the general behaviour of $R_{ei}$ is similar to the others. It is obvious that by decreasing the temperature (increasing $\bar{\beta}$) EFR of all interactions grows up. Also, in small values of collision energy, the effective potential is more impressive and the EFR decreases via increasing the collision energy.
Figure (\ref{fig:fig7}) shows the EFR of two-temperature plasma with respect to electron and ion temperatures for electron-electron and ion-ion interactions with fixed values of collision energy and plasma density. Here, both parameters, i.e. $R_{ee}$ and $R_{ii}$, have completely the same behavior. Decreasing the temperatures takes the system to quantum regime and the EFR grows up. In order to show the roles of electron and ion temperatures more clearly, the EFR of electron-ion interaction with respect to temperatures has been represented in Fig. (\ref{fig:fig8}) with contour plot. It is obvious that in very high temperature (classical limit for the temperature of both species) EFR vanishes. However, EFR is more sensitive to electron temperature in comparison with ion temperature. By decreasing the electron temperature EFR starts to grow.

\section{conclusions}\label{5}	
Recently, entanglement of quantum states is investigated in different areas of physics such as spin chains in condensed matter, non inertial frames in relativist quantum information, entanglement generation due to expansion of universe in cosmology and Lorentz violation in high energy physics. Plasmas are highly interacting environment and specially in quantum regime the entanglement of an incident particle and the constituents of plasma can be affected by an effective potential.

The EFR of different effective potentials in different kinds of plasmas has been considered. We considered a non ideal dens plasma and investigated the influence of effective potential on the EFR. Our consideration is restricted to special circumstances in order to obtain analytical relationships. Two different regions which satisfy the circumstances have been investigated.

We showed that in region $\mathrm{\rom{1}}$ or low temperature limit, the EFR is independent of temperature and it is monotonically increasing function of the collision energy. Although, the EFR in regin $\mathrm{\rom{1}}$ does not change by temperature variations. However, in region $\mathrm{\rom{3}}$ or high temperature limit, an explicit dependency to temperature is evident and the EFR decreases by increasing temperature. We showed that only the EFR of electron-ion interaction vanishes in infinite temperature limit while it has non-zero value for electron-electron and ion-ion interactions.

For a two-temperature dens plasma model with fixed density and collision energy, decreasing both temperatures leads to growing of EFR for all kinds of interactions. Of course, the EFR of electron-ion interaction is more sensitive to electron temperature either to ion one.
	
\section{Acknowledgment}\label{6}
This work was partially supported by the Ferdowsi University of Mashhad under Grant No. 3/43953.	

\bibliography{refs}

\end{document}